\documentclass[reprint,aps,prb,showpacs,amsmath,amssymb]{revtex4-1}

\usepackage{color}
\usepackage{amsfonts}
\usepackage{amssymb}
\usepackage{amsthm}
\usepackage{graphicx}
\usepackage{verbatim}
\usepackage{algpseudocode}  % predefined set of commands for use with algorithmicx

\newcommand{\Eq}   [1]{Eq.~(\ref{#1})}

\newcommand{\trm}     {\textrm}

\newcommand{\rr}{\mathbf{r}}
\newcommand{\rp}{\mathbf{r}^\prime}
\newcommand{\R} {\mathbf{R}}
\newcommand{\rh}{\mathbf{r}_h}
\newcommand{\rg}{\mathbf{r}_g}

\newcommand{\brho}{\bar{\rho}}

\newcommand{\bv}{\bar{v}}

\newcommand{\coulombd}{\vartheta(\rr_h,\rr_g)}
\newcommand{\rcmax} {r^\trm{\scriptsize{max}}_c}
\newcommand{\integ} {\trm{{\scriptsize{int}}}}

\newcommand{\Ri} {\mathbf{R}_i}
\newcommand{\Rj} {\mathbf{R}_j}
\newcommand{\Rk} {\mathbf{R}_k}
\newcommand{\Rl} {\mathbf{R}_l}

\newcommand{\Rji} {\mathbf{R}_{ji}}
\newcommand{\Rij} {\mathbf{R}_{ij}}
\newcommand{\Rli} {\mathbf{R}_{li}}
\newcommand{\Rki} {\mathbf{R}_{ki}}
\newcommand{\Rai} {\mathbf{R}_{ai}}

\newcommand{\Fetch}{$\rhd$ fetch}

\newcommand{\Accu} {$\diamond$ accumulate}
\newcommand{\Calc} {$\rhd$ calculate}

\newcommand{\Eval} {$\rhd$ evaluate}
\newcommand{\Integ}{$\rhd$ integrate}

\newcommand{\Store}{$\rhd$ store}

\begin{document}

\title{Linear--scaling implementation of exact exchange\\
using localized numerical orbitals and contraction reduction integrals}

\author{Lionel A. Truflandier}
\email{l.truflandier@ism.u-bordeaux1.fr}
\affiliation{London Centre for Nanotechnology, UCL, 17-19 Gordon Street, London WC1H 0AH, UK}
\affiliation{Thomas Young Centre and Department of Physics \& Astronomy, UCL, Gower Street, London WC1E 6BT, UK}
\altaffiliation{Present address: Institut des Sciences Mol{\'e}culaires, Universit{\'e} Bordeaux I, 351 Cours de la Lib{\'e}ration, 33405 Talence, France}

\author{Tsuyoshi Miyazaki}
%\email{miyazaki.tsuyoshi@nims.go.jp}
\affiliation{National Institute for Materials Science, 1-2-1 Sengen, Tsukuba, Ibaraki 305-0045, Japan}

\author{David R. Bowler}
\email{david.bowler@ucl.ac.uk}
\affiliation{London Centre for Nanotechnology,
UCL, 17-19 Gordon Street, London WC1H 0AH, UK}
\affiliation{Thomas Young Centre and Department of Physics \& Astronomy, UCL, Gower Street, London WC1E 6BT, UK}
\affiliation{International Center for Materials Nanoarchitectonics, National Institute for Materials Science, 
1-1 Namiki, Tsukuba, Ibaraki 305-0044, Japan}

\date{\today}

\begin{abstract}
  We present enhancements to the computational
  efficiency of exact exchange calculations using the density matrix
  and local support functions. We introduce a numerical method which
  avoids the explicit calculation the four-center two-electron repulsion
  integrals and reduces the prefactor scaling by a factor $N$,
  where $N$ is the number of atoms within the range of the exact exchange
  Hamiltonian. This approach is based on a contraction-reduction
  scheme, and takes advantage of the discretization space which enables the 
  direct summation over the support functions in a localized space. Using the
  sparsity property of the density matrix, the scaling of the prefactor 
  can be further reduced to reach asymptotically $O(N)$.
\end{abstract}

\pacs{{71.15.Dx} {71.15.Ap} {71.15.Mb} {02.60.Jh}}

\maketitle 

The calculation of exchange energy as found in the ``Fock--exchange'' for Hartree--Fock (HF) theory 
or ``exact--exchange'' for Kohn--Sham (KS) density functional theory (DFT) is well-known to be time 
consuming, where for a naive implementation the scaling increases with the fourth power of the number 
of atoms, $N$, or the number of basis states. It has therefore been the focus of considerable 
efforts to improve both the efficiency and the scaling. Much of the work has taken place within the 
quantum chemistry community focussing on approaches using Gaussian-type orbitals (GTO) or exponential-type 
functions for the radial part such as the well-known Slater-type orbitals (STO). More recently, there has been 
interest in efficient implementation within the periodic density functional theory community, which traditionally 
use plane waves as basis functions (along with atomic pseudopotentials) and required discrete fast Fourier transform 
(FFT) technologies. At the same time, linear scaling, or $O(N)$, approaches to finding the electronic ground state have 
emerged over the last ten to fifteen years,\cite{ochsenfeld_linearscaling_2007,bowler_on_2012} and new schemes for the 
evaluation of exact exchange energy have to be developed including the specifications of the $O(N)$ techniques. We report 
a novel approach to improving the efficiency of exchange calculations for \emph{any} localised basis functions, which fits 
naturally within the formalism of linear scaling DFT calculations ; more specifically with the code 
\textsc{Conquest}\cite{Bowler:2002pt,Miyazaki:2004ee,Bowler:2006wa}. 

Within the framework of the HF theory, 
the exchange energy for a closed-shell system can be written as:
\begin{equation}
  \label{eq:1}
  E_{x} = -\frac{1}{4}\int d\rr d\rp \frac{\rho(\rr,\rp) \rho(\rp,\rr)}{\vert \rr - \rp \vert},
\end{equation}
using the definition of the density matrix in terms of the molecular eigenstates $\psi_{n}(\rr)$,
\begin{equation}
  \label{eq:2}
  \rho(\rr,\rp) = 2\sum_{n} \psi_{n}(\rp)\psi_{n}^{\star}(\rr),
\end{equation}
where $n$ runs over the doubly occupied orbitals. This yields a more explicit expression for
the exchange energy:
\begin{equation}
  \label{eq:3}
  E_{x} = -\sum_{nm}\int d\rr d\rp \frac{\psi^{\star}_{m}(\rr) \psi^{\star}_{n}(\rp) \psi_{n}(\rr)\psi_{m}(\rp)}{\vert \rr - \rp \vert}.
\end{equation}
The typical procedure in quantum chemistry is to express the exchange energy of~\Eq{eq:3} in terms of 4-center
2-electron repulsion integrals (ERI) by expanding $\psi_{n}(\rr)$ onto a linear combination of real atom-centered 
functions $\varphi_{i}(\rr)$. Using the standard notation, the ERI formally reads:
\begin{equation}
  \label{eq:4}
  (ik|lj) = \int d\rr d\rp\frac{\varphi_{i}(\rr)\varphi_{k}(\rr)\varphi_{l}(\rp)\varphi_{j}(\rp)}{\vert \rr - \rp \vert}
\end{equation}
The earliest approaches to linear scaling exchange used pre-screening on the integrals and the density matrix based on an assumed decay rate ;
see for instance the LinK\cite{ochsenfeld_linear_1998} and ONX\cite{schwegler_linear_1996,schwegler_linear_1997} algorithms. 
These analytical methods have been successfully applied in various quantum chemistry codes, though relying 
on specific basis sets such as Gaussian-type orbitals (GTO). Alternative efficient solutions generally make the use of a 3-center 
reduction scheme,\cite{polly_fast_2004,neese_efficient_2009,watson_density_2003,krykunov_hybrid_2009,friesner_correlated_2011}  
deriving from the density-fitting approach of Baerends and Roos for
Slater-type orbtials (STO),\cite{baerends_self-consistent_1973} 
and Dunlap and coworkers\cite{dunlap_approximations_1979,dunlap_first-row_1979} for GTO basis sets.
Different approaches with a similar spirit, such as the pseudo--spectral\cite{richard_a._solution_1985,friesner_solution_1986} 
or the resolution of identity methods\cite{eichkorn_auxiliary_1995,eichkorn_auxiliary_1997} have also demonstrated 
to be efficient, and improvements are still explored by many groups.\cite{friesner_correlated_2011,izsak_overlap_2011,reine_variational_2008}

The plane wave basis sets common within periodic DFT implementations make these approaches impossible, and most implementations 
concentrate on reciprocal space using FFTs.\cite{gygi_self-consistent_1986,chawla_exact_1998,paier_perdewburkeernzerhof_2005}
It should be mentioned here that acceptable accuracy is obtained only if an adequate treatment of the Coulomb singularities
are considered.\cite{Broqvist:2009kx,Duchemin:2010fk,Holzwarth:2011ys,Sorouri:2006uq,Spencer:2008vn}. Recently, using a transformation 
of the Kohn-Sham orbitals to maximally localised Wannier functions, a linear scaling calculation of the exchange potential has been 
demonstrated.\cite{wu_order-n_2009} Localised numerical orbital DFT approaches to exchange include the semi-analytic solution given 
by Toyoda and Ozaki\cite{toyoda_numerical_2009,toyoda_liberi:_2010} combining fast-spherical Bessel transform for the radial integration 
and a more traditional analytic method for the spherical harmonic part. A numerical scheme has also been proposed by Shang et 
al.\cite{shang_implementation_2010} where ERI are computed by solving numerically the Poisson's equation
%\begin{eqnarray}
%\label{eq:5}
%	  \nabla^2 v_{lj}(\rr) = -4\pi\rho_{lj}(\rr).
%\end{eqnarray} 
for each localized pair-density $\rho_{lj}=\varphi_l\varphi_j$, and integrating in real space.
%\begin{eqnarray}
%\label{eq:6}
%	  (ik|lj) = \int\rho_{ik}(\rr)v_{lj}(\rr)d\rr.
%\end{eqnarray}
Similarly to planewave periodic exchange calculations, the main drawback resides in the accuracy of the Poisson solver. All the methods 
outlined above involves the explicit calculation of the full or screened set of ERIs.

We introduce instead a route which circumvents the calculation of the four-center integrals and works for any smooth finite-range functions, which is particulary well suited for $O(N)$ approaches. In standard linear scaling theory, the density matrix 
is used as the fundamental variable and is written in a separable form in terms of localised orbitals, also called support functions
$\phi_{i}(\rr)$,
\begin{equation}
  \label{eq:7}
  \rho(\rr,\rp) = 2\sum_{ij}\phi_{i}(\rr) K_{ij} \phi_{j}(\rp)
\end{equation}
where $K_{ij}$ is the density matrix in the representation of the support functions, also known as the density kernel. Linear 
scaling is achieved when the support functions, centred on the atomic
positions $R_i$, are strictly localised in space and a cutoff is 
applied to $K_{ij}$ so that,
\begin{equation}
  \label{eq:8}
  K_{ij} = 0\ \trm{for}\ \vert R_{i} - R_{j}\vert > R_{K},
\end{equation}
with $R_K$ the density matrix range. From \Eq{eq:7} we can therefore write the exchange energy as:
\begin{eqnarray}
  \label{eq:9}
  E_{x} &=& -\sum_{ijkl} \int d\rr d\rp 
	    \frac{\phi_{i}(\rr) K_{ij} \phi_{j}(\rp) \phi_{k}(\rr) K_{kl} \phi_{l}(\rp)}{\vert \rr - \rp \vert}\\
  \label{eq:10}
	&=& -\sum_{ij} K_{ij} X_{ij}
\end{eqnarray}
with,
\begin{eqnarray}
  \label{eq:11}
  X_{ij}&=&  \sum_{kl} \int d\rr d\rp \frac{\rho_{ik}(\rr)K_{kl}\rho_{lj}(\rp)}{\vert \rr - \rp \vert}.
\end{eqnarray}
The exchange matrix $X$ becomes now the key quantity to calculate, and for $R_K\rightarrow\infty$, the resulting exchange
energy must be exact. We note that this form involves a contraction between $K$ and one set of local orbitals, seen as 
the sum over $l$ (or $k$) in \Eq{eq:11}. This type of contraction is frequently performed in \textsc{Conquest}\cite{Goringe:1997cy}, 
but in this case will reduce the prefactor for exchange energy calculation by removing one of the four centres of the ERI.
Moreover, if we apply a range $R_X$ to the exchange, such as
\begin{eqnarray}
  \label{eq:8bis}
  X_{ij} = 0\ \trm{for}\ \vert R_{i} - R_{j}\vert > R_{X},
\end{eqnarray}
we can then achieve linear scaling with the prefactor depending on the localisation of the matrix. The resulting method is 
not only efficient, but should be scalable in parallel, as it is compatible with the standard \textsc{Conquest} approach to 
matrix and support function operations. 

As mentioned in the introduction, the key part is to perform the sum over the index $l$ \emph{before} solving for the Coulomb potential 
of the pair densities; this simple re-ordering increases the efficiency of the procedure markedly, as we will show below.  
We define new contraction functions, $\Phi_{k}(\rp)$, as:
\begin{equation}
  \label{eq:12}
  \Phi_{k}(\rp) = \sum_{l} K_{kl}\phi_{l}(\rp)
\end{equation}
It should be outlined that the domain over which these functions are defined requires some care; this is detailed in the Appendix.  
The sum over $l$ need only include those support functions $\phi_{l}$ overlapping with $\phi_{j}$, as $\Phi_{k}$ will be multiplied by this 
function. Contracted densities are then defined as:
\begin{equation}
  \label{eq:13}
  \brho_{kj}(\rp) = \Phi_{k}(\rp)\phi_{j}(\rp),
\end{equation}
and the resulting Coulomb potential,
\begin{equation}
  \label{eq:14}
  \bv_{kj}(\rr) = \int d\rp \frac{\brho_{kj}(\rp)}{\vert \rr - \rp \vert},
\end{equation}
is calculated by solving Poisson's equation using, for instance, numerical FFT routines. As we discuss later, once the potential 
has been found a further contraction over $k$ is performed to create the function $\Omega_{j}(\rr)$, as:
\begin{equation}
  \label{eq:15}
  \Omega_{j}(\rr) = \sum_{k} \bv_{kj}(\rr) \phi_{k}(\rr),
\end{equation}
where, again, the sum over support functions $k$ need only include those functions which overlap with support function $i$.  
The matrix elements $X_{ij}$ are then calculated by integration:
\begin{equation}
  \label{eq:16}
  X_{ij} = \int d\rr \phi_{i}(\rr) \Omega_{j}(\rr).
\end{equation}
The set of function $\Omega_j$ is effectively defined by the density matrix range --normally applied to $K_{kl}$ 
in accordance with~\Eq{eq:8}-- and the need for $j$ to overlap with atoms $l$. There is therefore a clear route to efficient linear 
scaling exchange calculations within the standard approaches of $O(N)$ electronic structure codes. 

\begin{figure}[bht!]
  \centering
  \includegraphics[width=0.48\textwidth]{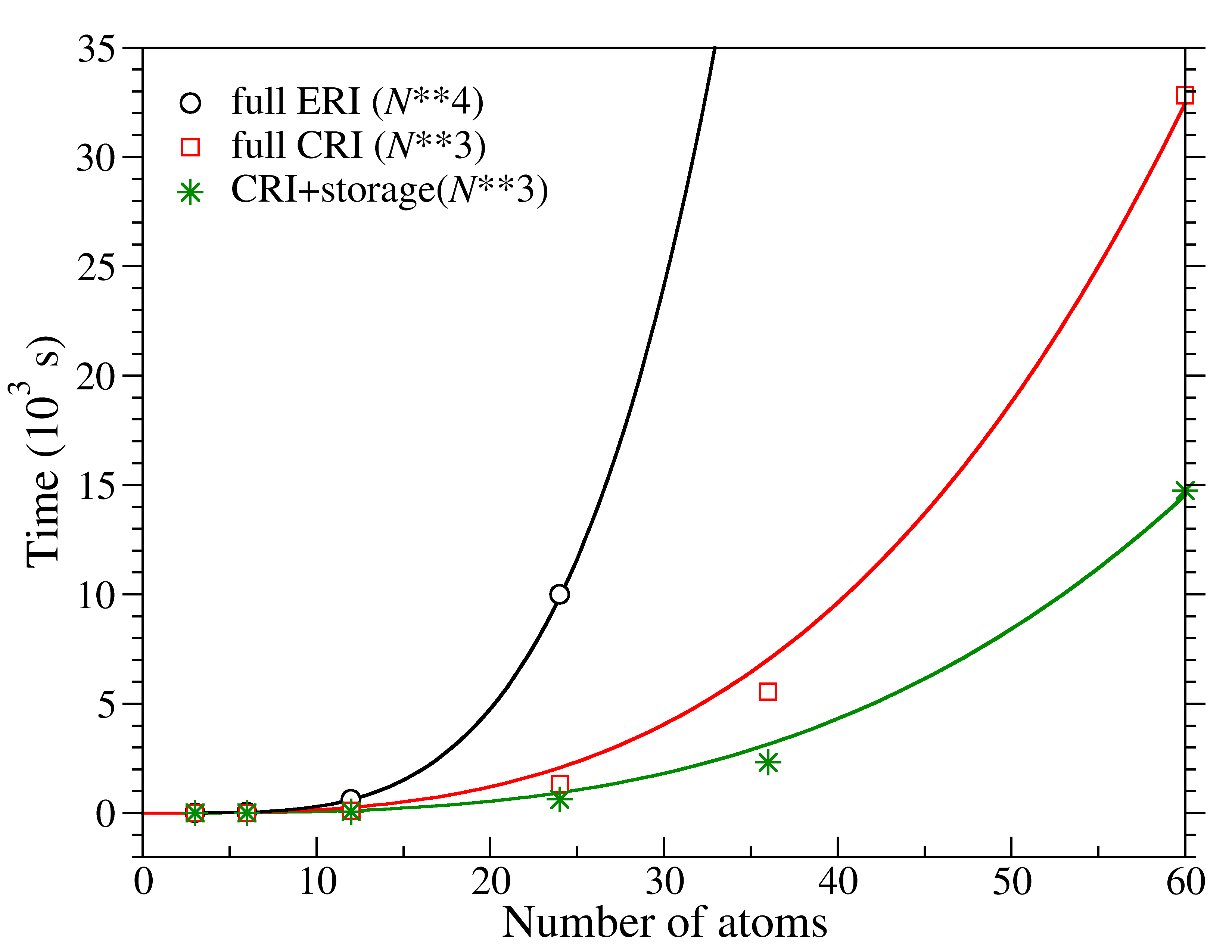}
  \caption{Comparison of CPU times necessary to compute EXX in 
	   isolated water clusters as a function of number atoms 
	   ($N$) using explicit ERI calculation and the CRI method. 
	   %Polynomial fits using $\alpha N^\beta$ are reported in 
	   %the parenthesis. 
	   Ideal $N^4$ and $N^3$ scalings are given by plain lines.
	   \label{fig:scaling1}}
\end{figure}
\begin{figure}[bht!]
  \centering
  \includegraphics[width=0.48\textwidth]{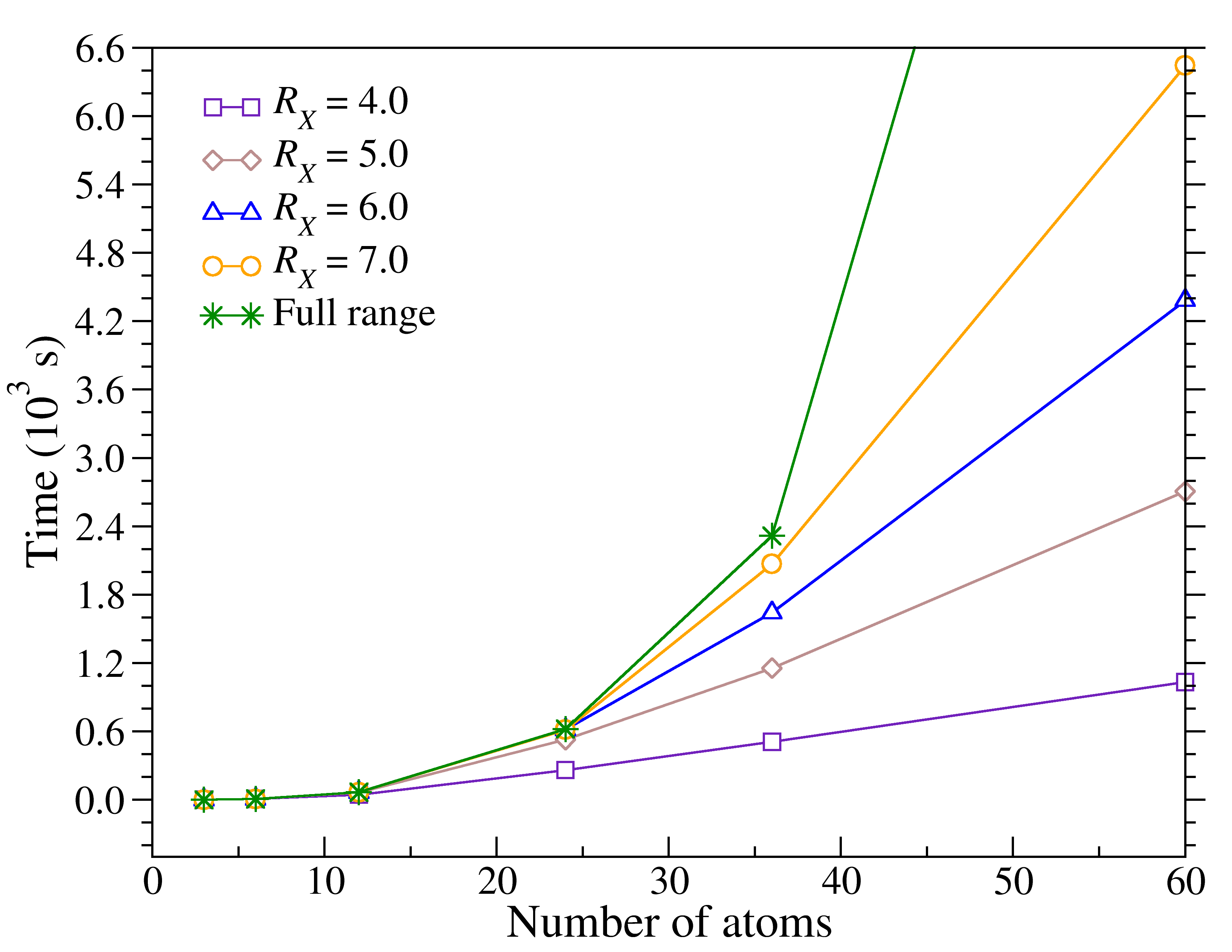}
  \caption{Variation of CPU time with respect to the range $R_X$ (in a.u.) for
	   the calculation of EXX in isolated water clusters using the CRI 
	   method.\label{fig:scaling2}}
\end{figure}

As previously remarked (see \Eq{eq:8bis}), the calculation time can be reduced further with a screening condition 
on exchange matrix elements $X_{ij}$. This is related to the sparsity property of $\rho(\rr,\rp),$\cite{kohn_density_1996} 
and the truncation of all the operators involved in the Hamiltonian.\cite{hernandez_linear-scaling_1996} From the algorithm of 
Fig.~\ref{fig:algoexx} (see Appendix), we note that the evaluation of the reduced potential $\bv_{kj}$ is performed within a 
3-index loop, which contrasts with the 4-index loop (over a limited
set of atoms) used for the
accumulation of temporary matrix $\Phi_k$. Another possibility 
would be to compute and store the set $\{\phi_l\}$ (and $\{\phi_k\}$) once, reducing formally --after the first cycle-- the 
execution time for the calculation of $\Phi_k$ to $N^3 (N^2)$ but increasing the data storage by $N (N^2)$, respectively.

Practical tests on the efficiency of the contraction-reduction integral (CRI) algorithm were performed on a set of 
isolated water clusters (H$_2$O)$_n$ ($n\leq20$) with fused cubes structures taken from the work of Wales and 
Hodges.\cite{wales_global_1998} Calculations of exchange energy were realized after the KS density matrix has been 
converged using the standard self-consistent-field (SCF) method. As a result, the timings presented below for exact 
exchange (EXX) energy can be compared to a single SCF cycle as found in HF or hybrid-DFT calculation. For this 
demonstration, single-$\zeta$ numerical pseudo-atomic orbitals\cite{sankey_ab_1989,junquera_numerical_2001,torralba_pseudo-atomic_2008} 
(NAO) have been used for hydrogen and oxygen with cutoff radii of 4.7 and 3.8 au, respectively. We emphasize
that the main conclusions of this work can be easily extended to more flexible basis sets, as far as the support 
functions are localized. SCF-KS and \emph{post}-EXX calculations were performed with a fixed grid spacing of 0.25 
au for the NAO discretization. This protocol allows us to realize fast enough computations on a single processor 
and also to draw qualitative conclusions on the exchange matrix range.

The central processing unit (CPU) times used for the computation of EXX are reported in Fig.~\ref{fig:scaling1}
as a function of the number of atoms, for various water clusters, using: (i) the explicit evaluation of the full set of ERI, 
(ii) the CRI approach, and (iii) the CRI approach with partial storage of the NAO during the construction of the temporary 
matrix $\Phi_k$ involved in the 4-index loop (see Appendix). Comparing the formal scalings obtained for the CRI methods against 
the full ERI approach, it becomes clear that the contraction-reduction
algorithm reduces the quartic scaling to to cubic scaling  
with respect to the size of the water clusters. Timings can be further reduced by requiring the storage of the NAOs 
(the set $\{\phi_l\}$). At this point we should emphasise that exchange energy values obtained with 
the three schemes are fully identical, their accuracies being only dependent on the Poisson solver used to evaluate the 
pair potential in \Eq{eq:14}.

Among the various numerical FFT-based methods, one can choose to evaluate the Coulomb potential in reciprocal or real space. 
Whereas the former is the most appropriate for periodic neutral systems --when the positively charged nuclei compensate exactly 
the electronic charge density-- it becomes less efficient for isolated and/or charged systems.\cite{castro_solution_2003} Several 
schemes have been developed to tackle this problem,\cite{jarvis_supercell_1997,martyna_reciprocal_1999,rozzi_exact_2006,dabo_electrostatics_2008}
Alternatives based on the discrete variable representation (DVR) of~\Eq{eq:14} which avoids the direct resolution of the Poisson 
equation have been proposed.\cite{lee_efficient_2008} The density is generally expanded in a direct product of one-dimensional 
localized real-space basis functions\cite{watson_linear-scaling_2008,lee_efficient_2008,varga_lagrange_2004} as for instance,
interpolating scaling functions (ISF). After extended comparisons between the DVR-ISF developed by Genovese et 
al.\cite{genovese_efficient_2006,genovese_efficient_2007} and corrected reciprocal FFT-based 
schemes,\cite{onida_ab_1995,blochl_electrostatic_1995,schultz_local_1999} we found that systematic 
convergence of the ERI is obtained with a better accuracy and at a lower cost using the real space Poisson solver.\cite{truflandier__2012}

\begin{figure}[tbh!]
  \centering
  \includegraphics[width=0.48\textwidth]{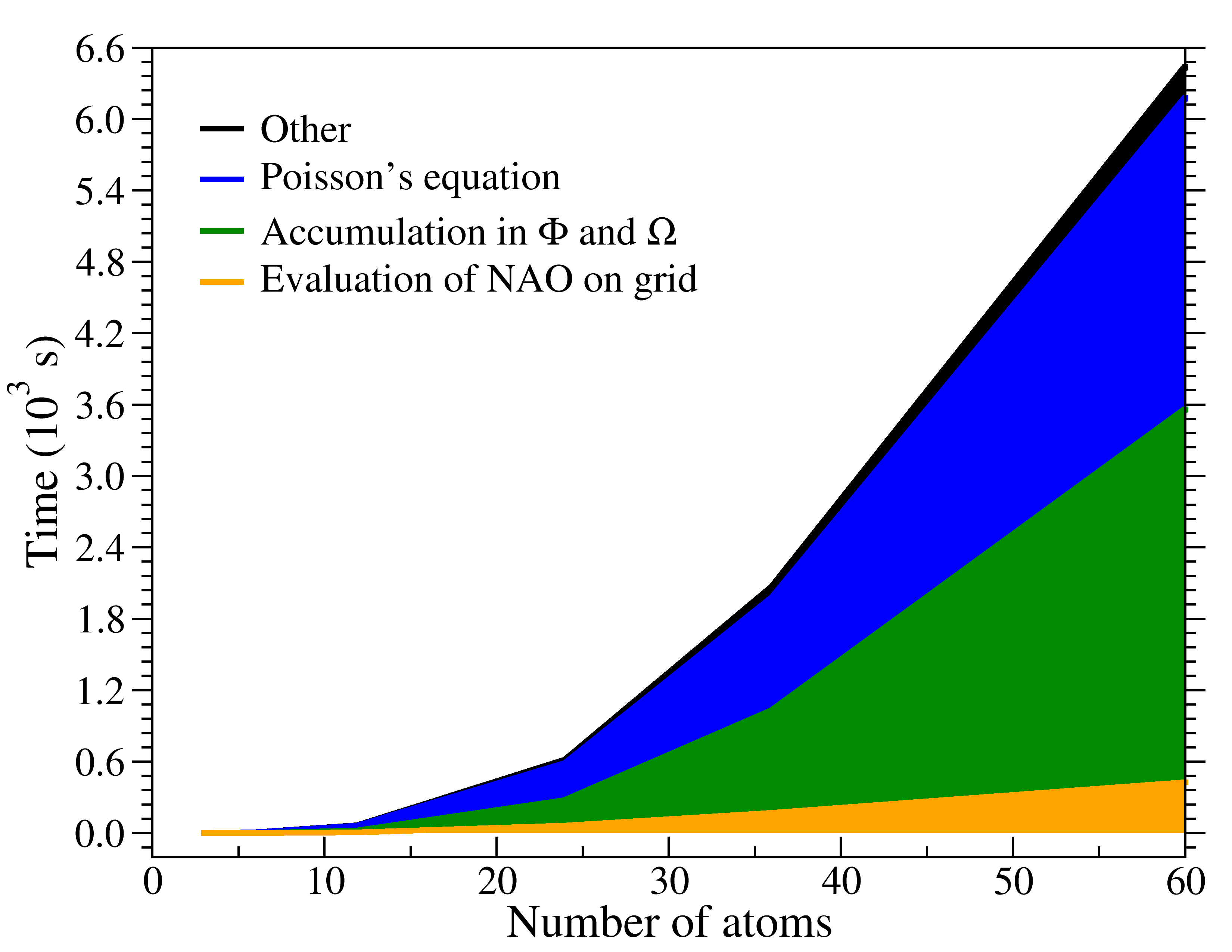}
  \caption{Cumulative decomposition of the total execution time for EXX
	   calculation in isolated water clusters using the CRI method along with
	   the storage of the NAO and a screening of $R_X=7.0$ au. Contributions of 
	   the three main routines are presented: the resolution of the Poisson equation, 
	   the accumulation in the temporary matrices $\Phi$ and $\Omega$, and the NAO 
	   discretization .\label{fig:scaling3}}
\end{figure}

As shown in Fig.~\ref{fig:scaling2}, if integral screenings is introduced within the CRI algorithm --see Fig.~\ref{fig:algoexx} 
of the Appendix-- the CPU time can be significantly reduced, allowing to reach the $O(N)$ regime for clusters with 
more than 36 atoms (at $R_X=7.0$ au). Computational ressources further decrease for shorter EXX range along with the 
faster observation of the linear-scaling regime. Figure~\ref{fig:scaling3} presents the decomposition of the total 
execution time involved the calculation of the EXX using the CRI approach for $R_X=7.0$ au. As it would be expected, 
most of the time is spent on the accumulation in the temporary matrices $\Phi_k$ and $\Omega_j$, and the Poisson solver. 
Evaluation of the support functions on the cubic grid do not impact to much on timings as far as the reduced overlap space 
technique is considered (see Appendix).

\begin{figure}[tbh!]
  \centering
  \includegraphics[width=0.48\textwidth]{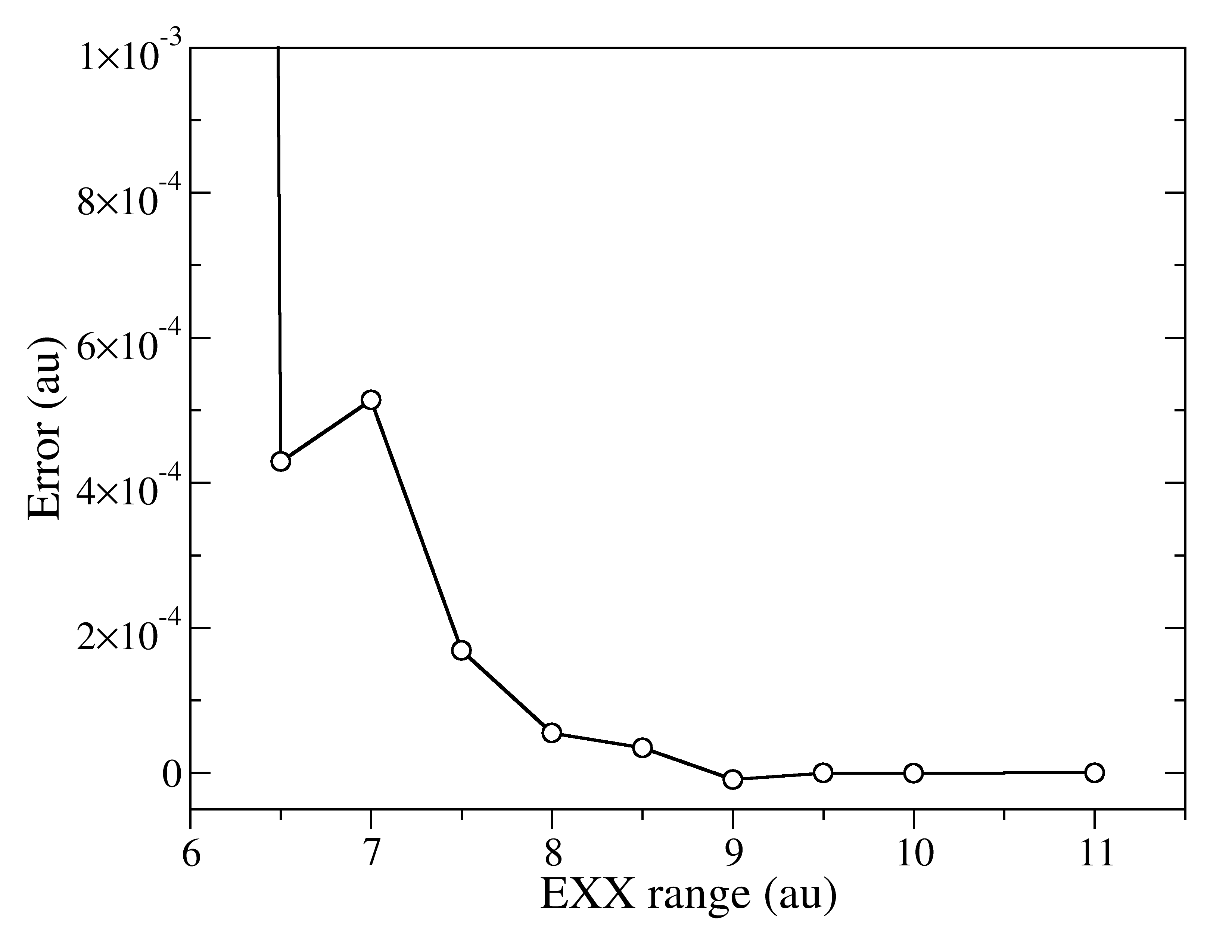}
  \caption{Convergence of the \textit{post}-EXX energy with respect to the exchange 
	   range $R_X$ for the cluster (H$_2$O)$_{20}$. Error is given with respect 
	   to the exact calculation.\label{fig:scaling4}}
\end{figure}

The \emph{post}-EXX accuracy with respect to $R_X$ is given in Fig.~\ref{fig:scaling4} for the cluster (H$_2$O)$_{20}$ using 
the ``boxkite'' structure,\cite{wales_global_1998} which is characterized by an edge length around 25.5 au. Because 
in the present study calculated EXX energy is not variational with respect to $R_X$, we do not expect a monotonic 
behavior for the plotted convergence profiles on Fig.~\ref{fig:scaling4}, where an accuracy below 1 mHa is found for $R_X\geq 6.5$. 
This can be compared to the density matrix convergence of 10$^{-7}$ Ha obtained at $R_K=5.0$ au. As a result, the price to paid for 
the fast computation of exchange energy using the CRI algorithm is the reduction of the accuracy, which in our case is acceptable 
considering the size of the system. It should be mentioned that the constant exchange cutoff used in the 3-index loop
of Fig.~\ref{fig:algoexx} can be different at the three stages. This
will allow acceleration of the convergence without significantly
affecting the efficiency.  This linear scaling approach will scale
in the same way as the other procedures in \textsc{Conquest}, opening
the way to efficient exact exchange calculations on 100,000+ atoms.

Finally, in this work we have shown that we are able to circumvent the $N^4$ scaling inferred by the standard calculation of
exchange without any approximation. Even if the non-local nature of the EXX interaction requires a larger range compared to
standard $O(N)$ DFT implementation, the linear-scaling regime is observable for a fair efficiency/accuracy ratio. In our case, 
computation time can be further reduced the fact that FFT-based Poisson' solvers are easily parallelizable along with a judicious
choice of the EXX matrix range.

\begin{acknowledgments}
L.A.T. was supported by the BBSRC grant BB/H024217/1, ``Linear Scaling Density Functional Theory for Biochemistry''.
The authors are grateful to L. Tong, M. J. Gillan and M. Toyoda for useful discussions.
\end{acknowledgments}
\clearpage
\appendix
\section{Implementation}
\label{sec:implementation}
To describe the practical implementation of the contraction-reduction integral algorithm we have to start from the explicit 
definition of the exchange matrix elements with respect to the ERIs and the basis set $\{\phi_{i}\}$. Within the discretized 
space \Eq{eq:11} can be written as:
\begin{eqnarray}
\nonumber
    X_{ij} = \sum_{kl}&&K_{kl}\sum_{hg}\phi_{i}(\rh-\Ri)\phi_{k}(\rh-\Rk)\coulombd \\
\label{eq:Wijdiscr}
    &&\times\phi_{l}(\rg-\Rl)\phi_{j}(\rg-\Rj)w(\rh)w(\rg),
\end{eqnarray}
where $\coulombd$ represents the 2-electron Coulomb operator. We made explicit in \Eq{eq:Wijdiscr} the fact that the 
support functions are centered on the nuclei positions $\{\Ri\}$.

The sets $\{w(\rr_h)\}$ and $\{w(\rr_g)\}$ account for the weight factors of the quadrature points
$\{\rr_h\}$ and $\{\rr_g\}$. We choose to work with an evenly spaced cubic grid where both $w(\rr_h)$ and
$w(\rr_g)$ simplify to $w_\integ = h^3_\integ$, with $h_\integ$ the grid spacing.
\footnote{There is no restriction for the generation of the sampling points and the corresponding weights.
It is well known that other nonlinear distributions allow to accelerate the convergence of numerical integration 
with respect to the size of the grid, mainly when we have to deal with singularities due to electronic cusps 
and nodal properties (for all--electron approaches) or Coulomb potentials. In this case, if one want to make 
an efficient use of FFT for the computation of Coulomb potential, we have to introduced transformation matrix 
to map the nonlinear distributions onto the FFT grid. We let this possibility open for future investigations.} 
Under the translation $\rr\rightarrow\rr+\Ri$, which leaves invariant the ERI, we obtain
\begin{eqnarray}
\nonumber
  X_{ij} = \sum_{kl}K_{kl}&&\sum_{hg}\phi_{i}(\rh)\phi_{k}(\rh-\Rki)\coulombd  \\
\label{eq:xijtransl}
  &&\times\phi_{l}(\rg-\Rli)\phi_{j}(\rg-\Rji)w^2_\integ, 
\end{eqnarray}
using $\Rij=\Ri-\Rj$. By virtue of the linearity of discretized space, we are allowed to introduce
the temporary matrix:
\begin{eqnarray}
\label{eq:phik}
  \Phi_{k}(\rg;\{\R_{li}\}) = \sum_{l}K_{kl}\phi_{l}(\rg-\Rli).
\end{eqnarray}
We emphasize that $\phi_{l}$ is evaluated on a cubic grid centered on the nucleus $i$.
The explicit expression for the reduced density of~\Eq{eq:13} is given by,
\begin{eqnarray}
\label{eq:rhokj}
  \brho_{kj}(\rg;\Rji,\{\Rli\}) = \Phi_{k}(\rg)\phi_{j}(\rg-\Rji).
\end{eqnarray}
The corresponding reduced pair potential $\bv_{lj}$ is obtained by solving the Poisson
equation. Finally, we introduce the temporary matrix:
\begin{eqnarray}
\nonumber
  \Omega_{j}(\rr_h;\Rji,\{\Rli\},\{\Rki&&\}) \\
\label{eq:omegaj}		
  = \sum_{k}&&\phi_k(\rr_h-\Rki)\bv_{kj}(\rh),
\end{eqnarray}
to perform the last numerical integration, yielding the exchange matrix elements:
\begin{eqnarray}
\label{eq:xij}				
  X_{ij}(\rr_h;\Rji,\{\Rli\},\{\Rki&&\})\nonumber\\
  = \sum_h&&\phi_i(\rr_h)\Omega_{j}(\rr_h)w_{\integ}.
\end{eqnarray}
We have made the dependence of the various matrix elements on
the translation vectors $\{\Rij\}$ explicit.
\begin{figure}[ht]
\caption{Algorithm describing exchange kernel formally scaling as $N^3$.}
%\begin{algorithm}
%\caption{}
\label{fig:algoexx}
\begin{algorithmic}[1]
%\sffamily
%\For{ atom pair $(i,j)$}
\Loop{ over atom $i$}
		\State\Eval\ and \Store\ $\phi_i$	
		\Loop{ over atom $j$}
				\If{$R_{ji} < R_{X}$}
						\State\Eval\ and \Store\ $\phi_j$
						\Loop{ over atom $k$}
								\If{$R_{ki} < R_{X}$}
										\State\Eval\ $\phi_k$ and \Store\ ?
										\Loop{ over atom $l$}
												\State\Fetch\ $K_{kl}$
												\If{$R_{li} < R_{X}$}
														\State\Eval\ $\phi_l$	and \Store\ ?
														\State\Accu\ $\Phi_k$
												\EndIf: $R_{X}$
										\EndLoop: $l$                        	
										\State\Calc\ $\brho_{kj}$		
										%\State\fbox{\Eval\ $\bv_{kj}$}
										\State\Eval\ $\bv_{kj}$
										%\State\Eval\ $\phi_k$	
										\State\Accu\ $\Omega_j$	
								\EndIf: $R_{X}$
						\EndLoop: $k$     
				\EndIf: $R_{X}$                   	
		\EndLoop: $j$
		%\State\Eval\ $\phi_i$	
		\State\Integ\ $X_{ij}$
\EndLoop: $i$                        		                        	
%\EndFor: $(i,j)$
\end{algorithmic}
%\end{algorithm}
\end{figure}
The CRI approach involves three main operations: (i) The projection of $\phi_i$ onto the discretized 
space, where both radial functions and spherical harmonics are evaluated on a cubic grid.
(ii) The summations of Eqs.~(\ref{eq:phik}) and (\ref{eq:omegaj}). (iii) The evaluation of
the pair potential $\bv_{kj}$.

The combination of local FFT grids\cite{skylaris_accurate_2001} with the locality property of the NAO easily fulfils the 
efficiency requirement. On each \textit{primary} atom $i$ a box is centered at the position $\R_i$. This box contains an ensemble 
of grid points called $\mathcal{B}_{i}$. For the NAO set $\{j,k,l\}$ in \Eq{eq:xijtransl} other boxes $\mathcal{B}_{a}$ are defined 
and translated along the vector $\Rai$. Here, we choose to work with identical cubic boxes of length $L\geq2\times\rcmax$, where $\rcmax$ 
is the largest confinement radius over the whole set of contracted support functions. Considering that the quadrature of \Eq{eq:xijtransl} 
is different from zero if significant overlap is deemed to exist between the orbital-pairs $ik$ and $lj$, we can first reduced the computational 
resources involved in (i) by defining reduced spaces as, 
\begin{eqnarray}
\label{eq:overlap}
  \mathcal{O}_{ab}=\mathcal{B}_{a}\cap\mathcal{B}_{b}
\end{eqnarray}
where $\mathcal{O}_{ab}$ is the overlap box of $\phi_{a}$ with $\phi_{b}$. Then the discretization of $\{\phi_{k},\phi_{l}\}$ 
is only realized for grid points common to the space span by $\phi_{i}$ and $\phi_{k}$, respectively. Secondly, by using the fact 
that the coordinate system is centered on the primary atom, we can introduced an efficient screening during the course of the 
calculation and reduced the computational time related to (ii) and (iii). Accumulation in the temporary matrices $\Phi_k$ and $\Omega_j$, 
which are centered on atom $i$, is performed if the distance $R_{ai}$ between the two distribution centers is below the EXX cutoff $R_{X}$.

%\bibliography{./bib/Friesner,./bib/Neese,./bib/Dunlap,./bib/Manby,./bib/Becke,./bib/Obara,./bib/Baerends,./bib/Barnett,
%./bib/Handy,./bib/FMM,./bib/Ahlrichs,./bib/STO,./bib/PW,./bib/CQreview,./bib/Scheffler,./bib/Shang,./bib/ISF,./bib/Wannier,
%./bib/Taleman,./bib/Toyoda,./bib/Beylkin,./bib/Sundholm,./bib/Tuckerman,./bib/Varga,./bib/Pulay,./bib/Kresse,./bib/ONETEP,
%./bib/DRBExtras,./bib/ONX,./bib/LinK,./bib/Compdetails}

%merlin.mbs apsrev4-1.bst 2010-07-25 4.21a (PWD, AO, DPC) hacked
%Control: key (0)
%Control: author (8) initials jnrlst
%Control: editor formatted (1) identically to author
%Control: production of article title (-1) disabled
%Control: page (0) single
%Control: year (1) truncated
%Control: production of eprint (0) enabled
%
%\bibliography{EXX_lioneltruflandier.bbl}

\begin{thebibliography}{57}%
\makeatletter
\providecommand \@ifxundefined [1]{%
 \@ifx{#1\undefined}
}%
\providecommand \@ifnum [1]{%
 \ifnum #1\expandafter \@firstoftwo
 \else \expandafter \@secondoftwo
 \fi
}%
\providecommand \@ifx [1]{%
 \ifx #1\expandafter \@firstoftwo
 \else \expandafter \@secondoftwo
 \fi
}%
\providecommand \natexlab [1]{#1}%
\providecommand \enquote  [1]{``#1''}%
\providecommand \bibnamefont  [1]{#1}%
\providecommand \bibfnamefont [1]{#1}%
\providecommand \citenamefont [1]{#1}%
\providecommand \href@noop [0]{\@secondoftwo}%
\providecommand \href [0]{\begingroup \@sanitize@url \@href}%
\providecommand \@href[1]{\@@startlink{#1}\@@href}%
\providecommand \@@href[1]{\endgroup#1\@@endlink}%
\providecommand \@sanitize@url [0]{\catcode `\\12\catcode `\$12\catcode
  `\&12\catcode `\#12\catcode `\^12\catcode `\_12\catcode `\%12\relax}%
\providecommand \@@startlink[1]{}%
\providecommand \@@endlink[0]{}%
\providecommand \url  [0]{\begingroup\@sanitize@url \@url }%
\providecommand \@url [1]{\endgroup\@href {#1}{\urlprefix }}%
\providecommand \urlprefix  [0]{URL }%
\providecommand \Eprint [0]{\href }%
\providecommand \doibase [0]{http://dx.doi.org/}%
\providecommand \selectlanguage [0]{\@gobble}%
\providecommand \bibinfo  [0]{\@secondoftwo}%
\providecommand \bibfield  [0]{\@secondoftwo}%
\providecommand \translation [1]{[#1]}%
\providecommand \BibitemOpen [0]{}%
\providecommand \bibitemStop [0]{}%
\providecommand \bibitemNoStop [0]{.\EOS\space}%
\providecommand \EOS [0]{\spacefactor3000\relax}%
\providecommand \BibitemShut  [1]{\csname bibitem#1\endcsname}%
\let\auto@bib@innerbib\@empty
%</preamble>
\bibitem [{\citenamefont {Ochsenfeld}\ \emph {et~al.}(2007)\citenamefont
  {Ochsenfeld}, \citenamefont {Kussmann},\ and\ \citenamefont
  {Lambrecht}}]{ochsenfeld_linearscaling_2007}%
  \BibitemOpen
  \bibfield  {author} {\bibinfo {author} {\bibfnamefont {C.}~\bibnamefont
  {Ochsenfeld}}, \bibinfo {author} {\bibfnamefont {J.}~\bibnamefont
  {Kussmann}}, \ and\ \bibinfo {author} {\bibfnamefont {D.~S.}\ \bibnamefont
  {Lambrecht}},\ }\href
  {http://onlinelibrary.wiley.com/doi/10.1002/9780470116449.ch1/summary}
  {\bibfield  {journal} {\bibinfo  {journal} {Reviews in Computational
  Chemistry}\ }\textbf {\bibinfo {volume} {23}},\ \bibinfo {pages} {1}
  (\bibinfo {year} {2007})}\BibitemShut {NoStop}%
\bibitem [{\citenamefont {Bowler}\ and\ \citenamefont
  {Miyazaki}(2012)}]{bowler_on_2012}%
  \BibitemOpen
  \bibfield  {author} {\bibinfo {author} {\bibfnamefont {D.~R.}\ \bibnamefont
  {Bowler}}\ and\ \bibinfo {author} {\bibfnamefont {T.}~\bibnamefont
  {Miyazaki}},\ }\href {\doibase 10.1088/0034-4885/75/3/036503} {\bibfield
  {journal} {\bibinfo  {journal} {Reports on Progress in Physics}\ }\textbf
  {\bibinfo {volume} {75}},\ \bibinfo {pages} {036503} (\bibinfo {year}
  {2012})}\BibitemShut {NoStop}%
\bibitem [{\citenamefont {Bowler}\ \emph {et~al.}(2002)\citenamefont {Bowler},
  \citenamefont {Miyazaki},\ and\ \citenamefont {Gillan}}]{Bowler:2002pt}%
  \BibitemOpen
  \bibfield  {author} {\bibinfo {author} {\bibfnamefont {D.~R.}\ \bibnamefont
  {Bowler}}, \bibinfo {author} {\bibfnamefont {T.}~\bibnamefont {Miyazaki}}, \
  and\ \bibinfo {author} {\bibfnamefont {M.~J.}\ \bibnamefont {Gillan}},\
  }\href@noop {} {\bibfield  {journal} {\bibinfo  {journal} {J. Phys.: Condens.
  Matter}\ }\textbf {\bibinfo {volume} {14}},\ \bibinfo {pages} {2781}
  (\bibinfo {year} {2002})}\BibitemShut {NoStop}%
\bibitem [{\citenamefont {Miyazaki}\ \emph {et~al.}(2004)\citenamefont
  {Miyazaki}, \citenamefont {Bowler}, \citenamefont {Choudhury},\ and\
  \citenamefont {Gillan}}]{Miyazaki:2004ee}%
  \BibitemOpen
  \bibfield  {author} {\bibinfo {author} {\bibfnamefont {T.}~\bibnamefont
  {Miyazaki}}, \bibinfo {author} {\bibfnamefont {D.~R.}\ \bibnamefont
  {Bowler}}, \bibinfo {author} {\bibfnamefont {R.}~\bibnamefont {Choudhury}}, \
  and\ \bibinfo {author} {\bibfnamefont {M.~J.}\ \bibnamefont {Gillan}},\
  }\href {\doibase 10.1063/1.1787832} {\bibfield  {journal} {\bibinfo
  {journal} {J. Chem. Phys.}\ }\textbf {\bibinfo {volume} {121}},\ \bibinfo
  {pages} {6186} (\bibinfo {year} {2004})}\BibitemShut {NoStop}%
\bibitem [{\citenamefont {Bowler}\ \emph {et~al.}(2006)\citenamefont {Bowler},
  \citenamefont {Choudhury}, \citenamefont {Gillan},\ and\ \citenamefont
  {Miyazaki}}]{Bowler:2006wa}%
  \BibitemOpen
  \bibfield  {author} {\bibinfo {author} {\bibfnamefont {D.~R.}\ \bibnamefont
  {Bowler}}, \bibinfo {author} {\bibfnamefont {R.}~\bibnamefont {Choudhury}},
  \bibinfo {author} {\bibfnamefont {M.~J.}\ \bibnamefont {Gillan}}, \ and\
  \bibinfo {author} {\bibfnamefont {T.}~\bibnamefont {Miyazaki}},\ }\href
  {http://dx.doi.org/10.1002/pssb.200541386} {\bibfield  {journal} {\bibinfo
  {journal} {phys. stat. sol. b}\ }\textbf {\bibinfo {volume} {243}},\ \bibinfo
  {pages} {989} (\bibinfo {year} {2006})},\ \bibinfo {note}
  {10.1002/pssb.200541386}\BibitemShut {NoStop}%
\bibitem [{\citenamefont {Ochsenfeld}\ \emph {et~al.}(1998)\citenamefont
  {Ochsenfeld}, \citenamefont {White},\ and\ \citenamefont
  {Head-Gordon}}]{ochsenfeld_linear_1998}%
  \BibitemOpen
  \bibfield  {author} {\bibinfo {author} {\bibfnamefont {C.}~\bibnamefont
  {Ochsenfeld}}, \bibinfo {author} {\bibfnamefont {C.~A.}\ \bibnamefont
  {White}}, \ and\ \bibinfo {author} {\bibfnamefont {M.}~\bibnamefont
  {Head-Gordon}},\ }\href {\doibase doi:10.1063/1.476741} {\bibfield  {journal}
  {\bibinfo  {journal} {The Journal of Chemical Physics}\ }\textbf {\bibinfo
  {volume} {109}},\ \bibinfo {pages} {1663} (\bibinfo {year}
  {1998})}\BibitemShut {NoStop}%
\bibitem [{\citenamefont {Schwegler}\ and\ \citenamefont
  {Challacombe}(1996)}]{schwegler_linear_1996}%
  \BibitemOpen
  \bibfield  {author} {\bibinfo {author} {\bibfnamefont {E.}~\bibnamefont
  {Schwegler}}\ and\ \bibinfo {author} {\bibfnamefont {M.}~\bibnamefont
  {Challacombe}},\ }\href {\doibase doi:10.1063/1.472135} {\bibfield  {journal}
  {\bibinfo  {journal} {The Journal of Chemical Physics}\ }\textbf {\bibinfo
  {volume} {105}},\ \bibinfo {pages} {2726} (\bibinfo {year}
  {1996})}\BibitemShut {NoStop}%
\bibitem [{\citenamefont {Schwegler}\ \emph {et~al.}(1997)\citenamefont
  {Schwegler}, \citenamefont {Challacombe},\ and\ \citenamefont
  {Head-Gordon}}]{schwegler_linear_1997}%
  \BibitemOpen
  \bibfield  {author} {\bibinfo {author} {\bibfnamefont {E.}~\bibnamefont
  {Schwegler}}, \bibinfo {author} {\bibfnamefont {M.}~\bibnamefont
  {Challacombe}}, \ and\ \bibinfo {author} {\bibfnamefont {M.}~\bibnamefont
  {Head-Gordon}},\ }\href {\doibase doi:10.1063/1.473833} {\bibfield  {journal}
  {\bibinfo  {journal} {The Journal of Chemical Physics}\ }\textbf {\bibinfo
  {volume} {106}},\ \bibinfo {pages} {9708} (\bibinfo {year}
  {1997})}\BibitemShut {NoStop}%
\bibitem [{\citenamefont {Polly}\ \emph {et~al.}(2004)\citenamefont {Polly},
  \citenamefont {Werner}, \citenamefont {Manby},\ and\ \citenamefont
  {Knowles}}]{polly_fast_2004}%
  \BibitemOpen
  \bibfield  {author} {\bibinfo {author} {\bibfnamefont {R.}~\bibnamefont
  {Polly}}, \bibinfo {author} {\bibfnamefont {H.}~\bibnamefont {Werner}},
  \bibinfo {author} {\bibfnamefont {F.~R.}\ \bibnamefont {Manby}}, \ and\
  \bibinfo {author} {\bibfnamefont {P.~J.}\ \bibnamefont {Knowles}},\ }\href
  {\doibase 10.1080/0026897042000274801} {\bibfield  {journal} {\bibinfo
  {journal} {Mol. Phys.}\ }\textbf {\bibinfo {volume} {102}},\ \bibinfo {pages}
  {2311} (\bibinfo {year} {2004})}\BibitemShut {NoStop}%
\bibitem [{\citenamefont {Neese}\ \emph {et~al.}(2009)\citenamefont {Neese},
  \citenamefont {Wennmohs}, \citenamefont {Hansen},\ and\ \citenamefont
  {Becker}}]{neese_efficient_2009}%
  \BibitemOpen
  \bibfield  {author} {\bibinfo {author} {\bibfnamefont {F.}~\bibnamefont
  {Neese}}, \bibinfo {author} {\bibfnamefont {F.}~\bibnamefont {Wennmohs}},
  \bibinfo {author} {\bibfnamefont {A.}~\bibnamefont {Hansen}}, \ and\ \bibinfo
  {author} {\bibfnamefont {U.}~\bibnamefont {Becker}},\ }\href {\doibase
  10.1016/j.chemphys.2008.10.036} {\bibfield  {journal} {\bibinfo  {journal}
  {Chem. Phys.}\ }\textbf {\bibinfo {volume} {356}},\ \bibinfo {pages} {98}
  (\bibinfo {year} {2009})}\BibitemShut {NoStop}%
\bibitem [{\citenamefont {Watson}\ \emph {et~al.}(2003)\citenamefont {Watson},
  \citenamefont {Handy},\ and\ \citenamefont {Cohen}}]{watson_density_2003}%
  \BibitemOpen
  \bibfield  {author} {\bibinfo {author} {\bibfnamefont {M.~A.}\ \bibnamefont
  {Watson}}, \bibinfo {author} {\bibfnamefont {N.~C.}\ \bibnamefont {Handy}}, \
  and\ \bibinfo {author} {\bibfnamefont {A.~J.}\ \bibnamefont {Cohen}},\ }\href
  {\doibase 10.1063/1.1604371} {\bibfield  {journal} {\bibinfo  {journal} {J.
  Chem. Phys.}\ }\textbf {\bibinfo {volume} {119}},\ \bibinfo {pages} {6475}
  (\bibinfo {year} {2003})}\BibitemShut {NoStop}%
\bibitem [{\citenamefont {Krykunov}\ \emph {et~al.}(2009)\citenamefont
  {Krykunov}, \citenamefont {Ziegler},\ and\ \citenamefont
  {Lenthe}}]{krykunov_hybrid_2009}%
  \BibitemOpen
  \bibfield  {author} {\bibinfo {author} {\bibfnamefont {M.}~\bibnamefont
  {Krykunov}}, \bibinfo {author} {\bibfnamefont {T.}~\bibnamefont {Ziegler}}, \
  and\ \bibinfo {author} {\bibfnamefont {E.~v.}\ \bibnamefont {Lenthe}},\
  }\href {\doibase 10.1002/qua.21985} {\bibfield  {journal} {\bibinfo
  {journal} {Int. J. Quantum Chem.}\ }\textbf {\bibinfo {volume} {109}},\
  \bibinfo {pages} {1676} (\bibinfo {year} {2009})}\BibitemShut {NoStop}%
\bibitem [{\citenamefont {Friesner}\ \emph {et~al.}(2011)\citenamefont
  {Friesner}, \citenamefont {Murphy}, \citenamefont {Beachy}, \citenamefont
  {Ringnalda}, \citenamefont {Pollard}, \citenamefont {Dunietz},\ and\
  \citenamefont {Cao}}]{friesner_correlated_2011}%
  \BibitemOpen
  \bibfield  {author} {\bibinfo {author} {\bibfnamefont {R.~A.}\ \bibnamefont
  {Friesner}}, \bibinfo {author} {\bibfnamefont {R.~B.}\ \bibnamefont
  {Murphy}}, \bibinfo {author} {\bibfnamefont {M.~D.}\ \bibnamefont {Beachy}},
  \bibinfo {author} {\bibfnamefont {M.~N.}\ \bibnamefont {Ringnalda}}, \bibinfo
  {author} {\bibfnamefont {W.~T.}\ \bibnamefont {Pollard}}, \bibinfo {author}
  {\bibfnamefont {B.~D.}\ \bibnamefont {Dunietz}}, \ and\ \bibinfo {author}
  {\bibfnamefont {Y.}~\bibnamefont {Cao}},\ }\href {\doibase doi:
  10.1021/jp9825157} {\bibfield  {journal} {\bibinfo  {journal} {J. Phys. Chem.
  A}\ }\textbf {\bibinfo {volume} {103}},\ \bibinfo {pages} {1913} (\bibinfo
  {year} {2011})}\BibitemShut {NoStop}%
\bibitem [{\citenamefont {Baerends}\ \emph {et~al.}(1973)\citenamefont
  {Baerends}, \citenamefont {Ellis},\ and\ \citenamefont
  {Ros}}]{baerends_self-consistent_1973}%
  \BibitemOpen
  \bibfield  {author} {\bibinfo {author} {\bibfnamefont {E.}~\bibnamefont
  {Baerends}}, \bibinfo {author} {\bibfnamefont {D.}~\bibnamefont {Ellis}}, \
  and\ \bibinfo {author} {\bibfnamefont {P.}~\bibnamefont {Ros}},\ }\href
  {\doibase 10.1016/0301-0104(73)80059-X} {\bibfield  {journal} {\bibinfo
  {journal} {Chem. Phys.}\ }\textbf {\bibinfo {volume} {2}},\ \bibinfo {pages}
  {41} (\bibinfo {year} {1973})}\BibitemShut {NoStop}%
\bibitem [{\citenamefont {Dunlap}\ \emph
  {et~al.}(1979{\natexlab{a}})\citenamefont {Dunlap}, \citenamefont
  {Connolly},\ and\ \citenamefont {Sabin}}]{dunlap_approximations_1979}%
  \BibitemOpen
  \bibfield  {author} {\bibinfo {author} {\bibfnamefont {B.~I.}\ \bibnamefont
  {Dunlap}}, \bibinfo {author} {\bibfnamefont {J.~W.~D.}\ \bibnamefont
  {Connolly}}, \ and\ \bibinfo {author} {\bibfnamefont {J.~R.}\ \bibnamefont
  {Sabin}},\ }\href {\doibase 10.1063/1.438728} {\bibfield  {journal} {\bibinfo
   {journal} {J. Chem. Phys.}\ }\textbf {\bibinfo {volume} {71}},\ \bibinfo
  {pages} {3396} (\bibinfo {year} {1979}{\natexlab{a}})}\BibitemShut {NoStop}%
\bibitem [{\citenamefont {Dunlap}\ \emph
  {et~al.}(1979{\natexlab{b}})\citenamefont {Dunlap}, \citenamefont
  {Connolly},\ and\ \citenamefont {Sabin}}]{dunlap_first-row_1979}%
  \BibitemOpen
  \bibfield  {author} {\bibinfo {author} {\bibfnamefont {B.~I.}\ \bibnamefont
  {Dunlap}}, \bibinfo {author} {\bibfnamefont {J.~W.~D.}\ \bibnamefont
  {Connolly}}, \ and\ \bibinfo {author} {\bibfnamefont {J.~R.}\ \bibnamefont
  {Sabin}},\ }\href {\doibase 10.1063/1.438313} {\bibfield  {journal} {\bibinfo
   {journal} {J. Chem. Phys.}\ }\textbf {\bibinfo {volume} {71}},\ \bibinfo
  {pages} {4993} (\bibinfo {year} {1979}{\natexlab{b}})}\BibitemShut {NoStop}%
\bibitem [{\citenamefont {Richard~A.}(1985)}]{richard_a._solution_1985}%
  \BibitemOpen
  \bibfield  {author} {\bibinfo {author} {\bibfnamefont {F.}~\bibnamefont
  {Richard~A.}},\ }\href {\doibase 10.1016/0009-2614(85)80121-4} {\bibfield
  {journal} {\bibinfo  {journal} {Chem. Phys. Lett.}\ }\textbf {\bibinfo
  {volume} {116}},\ \bibinfo {pages} {39} (\bibinfo {year} {1985})}\BibitemShut
  {NoStop}%
\bibitem [{\citenamefont {Friesner}(1986)}]{friesner_solution_1986}%
  \BibitemOpen
  \bibfield  {author} {\bibinfo {author} {\bibfnamefont {R.~A.}\ \bibnamefont
  {Friesner}},\ }\href {\doibase 10.1063/1.451237} {\bibfield  {journal}
  {\bibinfo  {journal} {J. Chem. Phys.}\ }\textbf {\bibinfo {volume} {85}},\
  \bibinfo {pages} {1462} (\bibinfo {year} {1986})}\BibitemShut {NoStop}%
\bibitem [{\citenamefont {Eichkorn}\ \emph {et~al.}(1995)\citenamefont
  {Eichkorn}, \citenamefont {Treutler}, \citenamefont {Öhm}, \citenamefont
  {Häser},\ and\ \citenamefont {Ahlrichs}}]{eichkorn_auxiliary_1995}%
  \BibitemOpen
  \bibfield  {author} {\bibinfo {author} {\bibfnamefont {K.}~\bibnamefont
  {Eichkorn}}, \bibinfo {author} {\bibfnamefont {O.}~\bibnamefont {Treutler}},
  \bibinfo {author} {\bibfnamefont {H.}~\bibnamefont {Öhm}}, \bibinfo {author}
  {\bibfnamefont {M.}~\bibnamefont {Häser}}, \ and\ \bibinfo {author}
  {\bibfnamefont {R.}~\bibnamefont {Ahlrichs}},\ }\href {\doibase
  10.1016/0009-2614(95)00838-U} {\bibfield  {journal} {\bibinfo  {journal}
  {Chem. Phys. Lett.}\ }\textbf {\bibinfo {volume} {242}},\ \bibinfo {pages}
  {652} (\bibinfo {year} {1995})}\BibitemShut {NoStop}%
\bibitem [{\citenamefont {Eichkorn}\ \emph {et~al.}(1997)\citenamefont
  {Eichkorn}, \citenamefont {Weigend}, \citenamefont {Treutler},\ and\
  \citenamefont {Ahlrichs}}]{eichkorn_auxiliary_1997}%
  \BibitemOpen
  \bibfield  {author} {\bibinfo {author} {\bibfnamefont {K.}~\bibnamefont
  {Eichkorn}}, \bibinfo {author} {\bibfnamefont {F.}~\bibnamefont {Weigend}},
  \bibinfo {author} {\bibfnamefont {O.}~\bibnamefont {Treutler}}, \ and\
  \bibinfo {author} {\bibfnamefont {R.}~\bibnamefont {Ahlrichs}},\ }\href
  {\doibase 10.1007/s002140050244} {\bibfield  {journal} {\bibinfo  {journal}
  {Theor. Chem. Acc.}\ }\textbf {\bibinfo {volume} {97}},\ \bibinfo {pages}
  {119} (\bibinfo {year} {1997})}\BibitemShut {NoStop}%
\bibitem [{\citenamefont {Izsák}\ and\ \citenamefont
  {Neese}(2011)}]{izsak_overlap_2011}%
  \BibitemOpen
  \bibfield  {author} {\bibinfo {author} {\bibfnamefont {R.}~\bibnamefont
  {Izsák}}\ and\ \bibinfo {author} {\bibfnamefont {F.}~\bibnamefont
  {Neese}},\ }\href {\doibase 10.1063/1.3646921} {\bibfield  {journal}
  {\bibinfo  {journal} {J. Chem. Phys.}\ }\textbf {\bibinfo {volume} {135}},\
  \bibinfo {pages} {144105} (\bibinfo {year} {2011})}\BibitemShut {NoStop}%
\bibitem [{\citenamefont {Reine}\ \emph {et~al.}(2008)\citenamefont {Reine},
  \citenamefont {Tellgren}, \citenamefont {Krapp}, \citenamefont {Kjærgaard},
  \citenamefont {Helgaker}, \citenamefont {Jansik}, \citenamefont {Ho̸st},\
  and\ \citenamefont {Salek}}]{reine_variational_2008}%
  \BibitemOpen
  \bibfield  {author} {\bibinfo {author} {\bibfnamefont {S.}~\bibnamefont
  {Reine}}, \bibinfo {author} {\bibfnamefont {E.}~\bibnamefont {Tellgren}},
  \bibinfo {author} {\bibfnamefont {A.}~\bibnamefont {Krapp}}, \bibinfo
  {author} {\bibfnamefont {T.}~\bibnamefont {Kjærgaard}}, \bibinfo {author}
  {\bibfnamefont {T.}~\bibnamefont {Helgaker}}, \bibinfo {author}
  {\bibfnamefont {B.}~\bibnamefont {Jansik}}, \bibinfo {author} {\bibfnamefont
  {S.}~\bibnamefont {Ho̸st}}, \ and\ \bibinfo {author} {\bibfnamefont
  {P.}~\bibnamefont {Salek}},\ }\href {\doibase 10.1063/1.2956507} {\bibfield
  {journal} {\bibinfo  {journal} {J. Chem. Phys.}\ }\textbf {\bibinfo {volume}
  {129}},\ \bibinfo {pages} {104101} (\bibinfo {year} {2008})}\BibitemShut
  {NoStop}%
\bibitem [{\citenamefont {Gygi}\ and\ \citenamefont
  {Baldereschi}(1986)}]{gygi_self-consistent_1986}%
  \BibitemOpen
  \bibfield  {author} {\bibinfo {author} {\bibfnamefont {F.}~\bibnamefont
  {Gygi}}\ and\ \bibinfo {author} {\bibfnamefont {A.}~\bibnamefont
  {Baldereschi}},\ }\href {\doibase 10.1103/PhysRevB.34.4405} {\bibfield
  {journal} {\bibinfo  {journal} {Phys. Rev. B}\ }\textbf {\bibinfo {volume}
  {34}},\ \bibinfo {pages} {4405} (\bibinfo {year} {1986})}\BibitemShut
  {NoStop}%
\bibitem [{\citenamefont {Chawla}\ and\ \citenamefont
  {Voth}(1998)}]{chawla_exact_1998}%
  \BibitemOpen
  \bibfield  {author} {\bibinfo {author} {\bibfnamefont {S.}~\bibnamefont
  {Chawla}}\ and\ \bibinfo {author} {\bibfnamefont {G.~A.}\ \bibnamefont
  {Voth}},\ }\href {\doibase 10.1063/1.476307} {\bibfield  {journal} {\bibinfo
  {journal} {J. Chem. Phys.}\ }\textbf {\bibinfo {volume} {108}},\ \bibinfo
  {pages} {4697} (\bibinfo {year} {1998})}\BibitemShut {NoStop}%
\bibitem [{\citenamefont {Paier}\ \emph {et~al.}(2005)\citenamefont {Paier},
  \citenamefont {Hirschl}, \citenamefont {Marsman},\ and\ \citenamefont
  {Kresse}}]{paier_perdewburkeernzerhof_2005}%
  \BibitemOpen
  \bibfield  {author} {\bibinfo {author} {\bibfnamefont {J.}~\bibnamefont
  {Paier}}, \bibinfo {author} {\bibfnamefont {R.}~\bibnamefont {Hirschl}},
  \bibinfo {author} {\bibfnamefont {M.}~\bibnamefont {Marsman}}, \ and\
  \bibinfo {author} {\bibfnamefont {G.}~\bibnamefont {Kresse}},\ }\href
  {\doibase 10.1063/1.1926272} {\bibfield  {journal} {\bibinfo  {journal} {J.
  Chem. Phys.}\ }\textbf {\bibinfo {volume} {122}},\ \bibinfo {pages} {234102}
  (\bibinfo {year} {2005})}\BibitemShut {NoStop}%
\bibitem [{\citenamefont {Broqvist}\ \emph {et~al.}(2009)\citenamefont
  {Broqvist}, \citenamefont {Alkauskas},\ and\ \citenamefont
  {Pasquarello}}]{Broqvist:2009kx}%
  \BibitemOpen
  \bibfield  {author} {\bibinfo {author} {\bibfnamefont {P.}~\bibnamefont
  {Broqvist}}, \bibinfo {author} {\bibfnamefont {A.}~\bibnamefont {Alkauskas}},
  \ and\ \bibinfo {author} {\bibfnamefont {A.}~\bibnamefont {Pasquarello}},\
  }\href {\doibase 10.1103/PhysRevB.80.085114} {\bibfield  {journal} {\bibinfo
  {journal} {Phys. Rev. B}\ }\textbf {\bibinfo {volume} {80}},\ \bibinfo
  {pages} {085114} (\bibinfo {year} {2009})}\BibitemShut {NoStop}%
\bibitem [{\citenamefont {Duchemin}\ and\ \citenamefont
  {Gygi}(2010)}]{Duchemin:2010fk}%
  \BibitemOpen
  \bibfield  {author} {\bibinfo {author} {\bibfnamefont {I.}~\bibnamefont
  {Duchemin}}\ and\ \bibinfo {author} {\bibfnamefont {F.}~\bibnamefont
  {Gygi}},\ }\href {\doibase 10.1016/j.cpc.2009.12.021} {\bibfield  {journal}
  {\bibinfo  {journal} {Comp. Phys. Commun.}\ }\textbf {\bibinfo {volume}
  {181}},\ \bibinfo {pages} {855 } (\bibinfo {year} {2010})}\BibitemShut
  {NoStop}%
\bibitem [{\citenamefont {Holzwarth}\ and\ \citenamefont
  {Xu}(2011)}]{Holzwarth:2011ys}%
  \BibitemOpen
  \bibfield  {author} {\bibinfo {author} {\bibfnamefont {N.~A.~W.}\
  \bibnamefont {Holzwarth}}\ and\ \bibinfo {author} {\bibfnamefont
  {X.}~\bibnamefont {Xu}},\ }\href {\doibase 10.1103/PhysRevB.84.113102}
  {\bibfield  {journal} {\bibinfo  {journal} {Phys. Rev. B}\ }\textbf {\bibinfo
  {volume} {84}},\ \bibinfo {pages} {113102} (\bibinfo {year}
  {2011})}\BibitemShut {NoStop}%
\bibitem [{\citenamefont {Sorouri}\ \emph {et~al.}(2006)\citenamefont
  {Sorouri}, \citenamefont {Foulkes},\ and\ \citenamefont
  {Hine}}]{Sorouri:2006uq}%
  \BibitemOpen
  \bibfield  {author} {\bibinfo {author} {\bibfnamefont {A.}~\bibnamefont
  {Sorouri}}, \bibinfo {author} {\bibfnamefont {W.~M.~C.}\ \bibnamefont
  {Foulkes}}, \ and\ \bibinfo {author} {\bibfnamefont {N.~D.~M.}\ \bibnamefont
  {Hine}},\ }\href {\doibase 10.1063/1.2166016} {\bibfield  {journal} {\bibinfo
   {journal} {J. Chem. Phys.}\ }\textbf {\bibinfo {volume} {124}},\ \bibinfo
  {eid} {064105} (\bibinfo {year} {2006})}\BibitemShut {NoStop}%
\bibitem [{\citenamefont {Spencer}\ and\ \citenamefont
  {Alavi}(2008)}]{Spencer:2008vn}%
  \BibitemOpen
  \bibfield  {author} {\bibinfo {author} {\bibfnamefont {J.}~\bibnamefont
  {Spencer}}\ and\ \bibinfo {author} {\bibfnamefont {A.}~\bibnamefont
  {Alavi}},\ }\href {\doibase 10.1103/PhysRevB.77.193110} {\bibfield  {journal}
  {\bibinfo  {journal} {Phys. Rev. B}\ }\textbf {\bibinfo {volume} {77}},\
  \bibinfo {pages} {193110} (\bibinfo {year} {2008})}\BibitemShut {NoStop}%
\bibitem [{\citenamefont {Wu}\ \emph {et~al.}(2009)\citenamefont {Wu},
  \citenamefont {Selloni},\ and\ \citenamefont {Car}}]{wu_order-n_2009}%
  \BibitemOpen
  \bibfield  {author} {\bibinfo {author} {\bibfnamefont {X.}~\bibnamefont
  {Wu}}, \bibinfo {author} {\bibfnamefont {A.}~\bibnamefont {Selloni}}, \ and\
  \bibinfo {author} {\bibfnamefont {R.}~\bibnamefont {Car}},\ }\href {\doibase
  10.1103/PhysRevB.79.085102} {\bibfield  {journal} {\bibinfo  {journal} {Phys.
  Rev. B}\ }\textbf {\bibinfo {volume} {79}},\ \bibinfo {pages} {085102}
  (\bibinfo {year} {2009})}\BibitemShut {NoStop}%
\bibitem [{\citenamefont {Toyoda}\ and\ \citenamefont
  {Ozaki}(2009)}]{toyoda_numerical_2009}%
  \BibitemOpen
  \bibfield  {author} {\bibinfo {author} {\bibfnamefont {M.}~\bibnamefont
  {Toyoda}}\ and\ \bibinfo {author} {\bibfnamefont {T.}~\bibnamefont {Ozaki}},\
  }\href {\doibase 10.1063/1.3082269} {\bibfield  {journal} {\bibinfo
  {journal} {J. Chem. Phys.}\ }\textbf {\bibinfo {volume} {130}},\ \bibinfo
  {pages} {124114} (\bibinfo {year} {2009})}\BibitemShut {NoStop}%
\bibitem [{\citenamefont {Toyoda}\ and\ \citenamefont
  {Ozaki}(2010)}]{toyoda_liberi:_2010}%
  \BibitemOpen
  \bibfield  {author} {\bibinfo {author} {\bibfnamefont {M.}~\bibnamefont
  {Toyoda}}\ and\ \bibinfo {author} {\bibfnamefont {T.}~\bibnamefont {Ozaki}},\
  }\href {\doibase 16/j.cpc.2010.03.019} {\bibfield  {journal} {\bibinfo
  {journal} {Comput. Phys. Comm.}\ }\textbf {\bibinfo {volume} {181}},\
  \bibinfo {pages} {1455} (\bibinfo {year} {2010})}\BibitemShut {NoStop}%
\bibitem [{\citenamefont {Shang}\ \emph {et~al.}(2010)\citenamefont {Shang},
  \citenamefont {Li},\ and\ \citenamefont {Yang}}]{shang_implementation_2010}%
  \BibitemOpen
  \bibfield  {author} {\bibinfo {author} {\bibfnamefont {H.}~\bibnamefont
  {Shang}}, \bibinfo {author} {\bibfnamefont {Z.}~\bibnamefont {Li}}, \ and\
  \bibinfo {author} {\bibfnamefont {J.}~\bibnamefont {Yang}},\ }\href {\doibase
  10.1021/jp908836z} {\bibfield  {journal} {\bibinfo  {journal} {J. Phys. Chem.
  A}\ }\textbf {\bibinfo {volume} {114}},\ \bibinfo {pages} {1039} (\bibinfo
  {year} {2010})}\BibitemShut {NoStop}%
\bibitem [{\citenamefont {Goringe}\ \emph {et~al.}(1997)\citenamefont
  {Goringe}, \citenamefont {Hern\'andez}, \citenamefont {Gillan},\ and\
  \citenamefont {Bush}}]{Goringe:1997cy}%
  \BibitemOpen
  \bibfield  {author} {\bibinfo {author} {\bibfnamefont {C.~M.}\ \bibnamefont
  {Goringe}}, \bibinfo {author} {\bibfnamefont {E.}~\bibnamefont
  {Hern\'andez}}, \bibinfo {author} {\bibfnamefont {M.~J.}\ \bibnamefont
  {Gillan}}, \ and\ \bibinfo {author} {\bibfnamefont {I.~J.}\ \bibnamefont
  {Bush}},\ }\href {http://dx.doi.org/10.1016/S0010-4655(97)00029-5} {\bibfield
   {journal} {\bibinfo  {journal} {Comp. Phys. Commun.}\ }\textbf {\bibinfo
  {volume} {102}},\ \bibinfo {pages} {1} (\bibinfo {year} {1997})},\ \bibinfo
  {note} {10.1016/S0010-4655(97)00029-5}\BibitemShut {NoStop}%
\bibitem [{\citenamefont {Wales}\ and\ \citenamefont
  {Hodges}(1998)}]{wales_global_1998}%
  \BibitemOpen
  \bibfield  {author} {\bibinfo {author} {\bibfnamefont {D.~J.}\ \bibnamefont
  {Wales}}\ and\ \bibinfo {author} {\bibfnamefont {M.~P.}\ \bibnamefont
  {Hodges}},\ }\href {\doibase 10.1016/S0009-2614(98)00065-7} {\bibfield
  {journal} {\bibinfo  {journal} {Chemical Physics Letters}\ }\textbf {\bibinfo
  {volume} {286}},\ \bibinfo {pages} {65} (\bibinfo {year} {1998})}\BibitemShut
  {NoStop}%
\bibitem [{\citenamefont {Sankey}\ and\ \citenamefont
  {Niklewski}(1989)}]{sankey_ab_1989}%
  \BibitemOpen
  \bibfield  {author} {\bibinfo {author} {\bibfnamefont {O.~F.}\ \bibnamefont
  {Sankey}}\ and\ \bibinfo {author} {\bibfnamefont {D.~J.}\ \bibnamefont
  {Niklewski}},\ }\href {\doibase 10.1103/PhysRevB.40.3979} {\bibfield
  {journal} {\bibinfo  {journal} {Phys. Rev. B}\ }\textbf {\bibinfo {volume}
  {40}},\ \bibinfo {pages} {3979} (\bibinfo {year} {1989})}\BibitemShut
  {NoStop}%
\bibitem [{\citenamefont {Junquera}\ \emph {et~al.}(2001)\citenamefont
  {Junquera}, \citenamefont {Paz}, \citenamefont {S{\'a}nchez-Portal},\ and\
  \citenamefont {Artacho}}]{junquera_numerical_2001}%
  \BibitemOpen
  \bibfield  {author} {\bibinfo {author} {\bibfnamefont {J.}~\bibnamefont
  {Junquera}}, \bibinfo {author} {\bibfnamefont {{\'O}.}~\bibnamefont {Paz}},
  \bibinfo {author} {\bibfnamefont {D.}~\bibnamefont {S{\'a}nchez-Portal}}, \
  and\ \bibinfo {author} {\bibfnamefont {E.}~\bibnamefont {Artacho}},\ }\href
  {\doibase 10.1103/PhysRevB.64.235111} {\bibfield  {journal} {\bibinfo
  {journal} {Physical Review B}\ }\textbf {\bibinfo {volume} {64}},\ \bibinfo
  {pages} {235111} (\bibinfo {year} {2001})}\BibitemShut {NoStop}%
\bibitem [{\citenamefont {Torralba}\ \emph {et~al.}(2008)\citenamefont
  {Torralba}, \citenamefont {Todorovi{\'c}}, \citenamefont {Br{\'a}zdov{\'a}},
  \citenamefont {Choudhury}, \citenamefont {Miyazaki}, \citenamefont {Gillan},\
  and\ \citenamefont {Bowler}}]{torralba_pseudo-atomic_2008}%
  \BibitemOpen
  \bibfield  {author} {\bibinfo {author} {\bibfnamefont {A.~S.}\ \bibnamefont
  {Torralba}}, \bibinfo {author} {\bibfnamefont {M.}~\bibnamefont
  {Todorovi{\'c}}}, \bibinfo {author} {\bibfnamefont {V.}~\bibnamefont
  {Br{\'a}zdov{\'a}}}, \bibinfo {author} {\bibfnamefont {R.}~\bibnamefont
  {Choudhury}}, \bibinfo {author} {\bibfnamefont {T.}~\bibnamefont {Miyazaki}},
  \bibinfo {author} {\bibfnamefont {M.~J.}\ \bibnamefont {Gillan}}, \ and\
  \bibinfo {author} {\bibfnamefont {D.~R.}\ \bibnamefont {Bowler}},\ }\href
  {\doibase 10.1088/0953-8984/20/29/294206} {\bibfield  {journal} {\bibinfo
  {journal} {Journal of Physics: Condensed Matter}\ }\textbf {\bibinfo {volume}
  {20}},\ \bibinfo {pages} {294206} (\bibinfo {year} {2008})}\BibitemShut
  {NoStop}%
\bibitem [{\citenamefont {Castro}\ \emph {et~al.}(2003)\citenamefont {Castro},
  \citenamefont {Rubio},\ and\ \citenamefont {Stott}}]{castro_solution_2003}%
  \BibitemOpen
  \bibfield  {author} {\bibinfo {author} {\bibfnamefont {A.}~\bibnamefont
  {Castro}}, \bibinfo {author} {\bibfnamefont {A.}~\bibnamefont {Rubio}}, \
  and\ \bibinfo {author} {\bibfnamefont {M.}~\bibnamefont {Stott}},\
  }\href@noop {} {\bibfield  {journal} {\bibinfo  {journal} {Can. J. Phys.}\
  }\textbf {\bibinfo {volume} {81}},\ \bibinfo {pages} {1151} (\bibinfo {year}
  {2003})}\BibitemShut {NoStop}%
\bibitem [{\citenamefont {Jarvis}\ \emph {et~al.}(1997)\citenamefont {Jarvis},
  \citenamefont {White}, \citenamefont {Godby},\ and\ \citenamefont
  {Payne}}]{jarvis_supercell_1997}%
  \BibitemOpen
  \bibfield  {author} {\bibinfo {author} {\bibfnamefont {M.~R.}\ \bibnamefont
  {Jarvis}}, \bibinfo {author} {\bibfnamefont {I.~D.}\ \bibnamefont {White}},
  \bibinfo {author} {\bibfnamefont {R.~W.}\ \bibnamefont {Godby}}, \ and\
  \bibinfo {author} {\bibfnamefont {M.~C.}\ \bibnamefont {Payne}},\ }\href
  {\doibase 10.1103/PhysRevB.56.14972} {\bibfield  {journal} {\bibinfo
  {journal} {Phys. Rev. B}\ }\textbf {\bibinfo {volume} {56}},\ \bibinfo
  {pages} {14972} (\bibinfo {year} {1997})}\BibitemShut {NoStop}%
\bibitem [{\citenamefont {Martyna}\ and\ \citenamefont
  {Tuckerman}(1999)}]{martyna_reciprocal_1999}%
  \BibitemOpen
  \bibfield  {author} {\bibinfo {author} {\bibfnamefont {G.~J.}\ \bibnamefont
  {Martyna}}\ and\ \bibinfo {author} {\bibfnamefont {M.~E.}\ \bibnamefont
  {Tuckerman}},\ }\href {\doibase 10.1063/1.477923} {\bibfield  {journal}
  {\bibinfo  {journal} {J. Chem. Phys.}\ }\textbf {\bibinfo {volume} {110}},\
  \bibinfo {pages} {2810} (\bibinfo {year} {1999})}\BibitemShut {NoStop}%
\bibitem [{\citenamefont {Rozzi}\ \emph {et~al.}(2006)\citenamefont {Rozzi},
  \citenamefont {Varsano}, \citenamefont {Marini}, \citenamefont {Gross},\ and\
  \citenamefont {Rubio}}]{rozzi_exact_2006}%
  \BibitemOpen
  \bibfield  {author} {\bibinfo {author} {\bibfnamefont {C.~A.}\ \bibnamefont
  {Rozzi}}, \bibinfo {author} {\bibfnamefont {D.}~\bibnamefont {Varsano}},
  \bibinfo {author} {\bibfnamefont {A.}~\bibnamefont {Marini}}, \bibinfo
  {author} {\bibfnamefont {E.~K.~U.}\ \bibnamefont {Gross}}, \ and\ \bibinfo
  {author} {\bibfnamefont {A.}~\bibnamefont {Rubio}},\ }\href {\doibase
  10.1103/PhysRevB.73.205119} {\bibfield  {journal} {\bibinfo  {journal} {Phys.
  Rev. B}\ }\textbf {\bibinfo {volume} {73}},\ \bibinfo {pages} {205119}
  (\bibinfo {year} {2006})}\BibitemShut {NoStop}%
\bibitem [{\citenamefont {Dabo}\ \emph {et~al.}(2008)\citenamefont {Dabo},
  \citenamefont {Kozinsky}, \citenamefont {{Singh-Miller}},\ and\ \citenamefont
  {Marzari}}]{dabo_electrostatics_2008}%
  \BibitemOpen
  \bibfield  {author} {\bibinfo {author} {\bibfnamefont {I.}~\bibnamefont
  {Dabo}}, \bibinfo {author} {\bibfnamefont {B.}~\bibnamefont {Kozinsky}},
  \bibinfo {author} {\bibfnamefont {N.~E.}\ \bibnamefont {{Singh-Miller}}}, \
  and\ \bibinfo {author} {\bibfnamefont {N.}~\bibnamefont {Marzari}},\ }\href
  {\doibase 10.1103/PhysRevB.77.115139} {\bibfield  {journal} {\bibinfo
  {journal} {Phys. Rev. B}\ }\textbf {\bibinfo {volume} {77}},\ \bibinfo
  {pages} {115139} (\bibinfo {year} {2008})}\BibitemShut {NoStop}%
\bibitem [{\citenamefont {Lee}\ and\ \citenamefont
  {Tuckerman}(2008)}]{lee_efficient_2008}%
  \BibitemOpen
  \bibfield  {author} {\bibinfo {author} {\bibfnamefont {H.}~\bibnamefont
  {Lee}}\ and\ \bibinfo {author} {\bibfnamefont {M.~E.}\ \bibnamefont
  {Tuckerman}},\ }\href {\doibase 10.1063/1.3036423} {\bibfield  {journal}
  {\bibinfo  {journal} {J. Chem. Phys.}\ }\textbf {\bibinfo {volume} {129}},\
  \bibinfo {pages} {224108} (\bibinfo {year} {2008})}\BibitemShut {NoStop}%
\bibitem [{\citenamefont {Watson}\ and\ \citenamefont
  {Hirao}(2008)}]{watson_linear-scaling_2008}%
  \BibitemOpen
  \bibfield  {author} {\bibinfo {author} {\bibfnamefont {M.~A.}\ \bibnamefont
  {Watson}}\ and\ \bibinfo {author} {\bibfnamefont {K.}~\bibnamefont {Hirao}},\
  }\href {\doibase 10.1063/1.3009264} {\ \textbf {\bibinfo {volume} {129}},\
  \bibinfo {pages} {184107} (\bibinfo {year} {2008})}\BibitemShut {NoStop}%
\bibitem [{\citenamefont {Varga}\ \emph {et~al.}(2004)\citenamefont {Varga},
  \citenamefont {Zhang},\ and\ \citenamefont
  {Pantelides}}]{varga_lagrange_2004}%
  \BibitemOpen
  \bibfield  {author} {\bibinfo {author} {\bibfnamefont {K.}~\bibnamefont
  {Varga}}, \bibinfo {author} {\bibfnamefont {Z.}~\bibnamefont {Zhang}}, \ and\
  \bibinfo {author} {\bibfnamefont {S.~T.}\ \bibnamefont {Pantelides}},\ }\href
  {\doibase 10.1103/PhysRevLett.93.176403} {\bibfield  {journal} {\bibinfo
  {journal} {Phys. Rev. Lett.}\ }\textbf {\bibinfo {volume} {93}},\ \bibinfo
  {pages} {176403} (\bibinfo {year} {2004})}\BibitemShut {NoStop}%
\bibitem [{\citenamefont {Genovese}\ \emph {et~al.}(2006)\citenamefont
  {Genovese}, \citenamefont {Deutsch}, \citenamefont {Neelov}, \citenamefont
  {Goedecker},\ and\ \citenamefont {Beylkin}}]{genovese_efficient_2006}%
  \BibitemOpen
  \bibfield  {author} {\bibinfo {author} {\bibfnamefont {L.}~\bibnamefont
  {Genovese}}, \bibinfo {author} {\bibfnamefont {T.}~\bibnamefont {Deutsch}},
  \bibinfo {author} {\bibfnamefont {A.}~\bibnamefont {Neelov}}, \bibinfo
  {author} {\bibfnamefont {S.}~\bibnamefont {Goedecker}}, \ and\ \bibinfo
  {author} {\bibfnamefont {G.}~\bibnamefont {Beylkin}},\ }\href {\doibase
  10.1063/1.2335442} {\bibfield  {journal} {\bibinfo  {journal} {J. Chem.
  Phys.}\ }\textbf {\bibinfo {volume} {125}},\ \bibinfo {pages} {074105}
  (\bibinfo {year} {2006})}\BibitemShut {NoStop}%
\bibitem [{\citenamefont {Genovese}\ \emph {et~al.}(2007)\citenamefont
  {Genovese}, \citenamefont {Deutsch},\ and\ \citenamefont
  {Goedecker}}]{genovese_efficient_2007}%
  \BibitemOpen
  \bibfield  {author} {\bibinfo {author} {\bibfnamefont {L.}~\bibnamefont
  {Genovese}}, \bibinfo {author} {\bibfnamefont {T.}~\bibnamefont {Deutsch}}, \
  and\ \bibinfo {author} {\bibfnamefont {S.}~\bibnamefont {Goedecker}},\ }\href
  {\doibase 10.1063/1.2754685} {\bibfield  {journal} {\bibinfo  {journal} {J.
  Chem. Phys.}\ }\textbf {\bibinfo {volume} {127}},\ \bibinfo {pages} {054704}
  (\bibinfo {year} {2007})}\BibitemShut {NoStop}%
\bibitem [{\citenamefont {Onida}\ \emph {et~al.}(1995)\citenamefont {Onida},
  \citenamefont {Reining}, \citenamefont {Godby}, \citenamefont {Del~Sole},\
  and\ \citenamefont {Andreoni}}]{onida_ab_1995}%
  \BibitemOpen
  \bibfield  {author} {\bibinfo {author} {\bibfnamefont {G.}~\bibnamefont
  {Onida}}, \bibinfo {author} {\bibfnamefont {L.}~\bibnamefont {Reining}},
  \bibinfo {author} {\bibfnamefont {R.~W.}\ \bibnamefont {Godby}}, \bibinfo
  {author} {\bibfnamefont {R.}~\bibnamefont {Del~Sole}}, \ and\ \bibinfo
  {author} {\bibfnamefont {W.}~\bibnamefont {Andreoni}},\ }\href {\doibase
  10.1103/PhysRevLett.75.818} {\bibfield  {journal} {\bibinfo  {journal} {Phys.
  Rev. Lett.}\ }\textbf {\bibinfo {volume} {75}},\ \bibinfo {pages} {818}
  (\bibinfo {year} {1995})}\BibitemShut {NoStop}%
\bibitem [{\citenamefont {Blöchl}(1995)}]{blochl_electrostatic_1995}%
  \BibitemOpen
  \bibfield  {author} {\bibinfo {author} {\bibfnamefont {P.~E.}\ \bibnamefont
  {Blöchl}},\ }\href {\doibase 10.1063/1.470314} {\bibfield  {journal}
  {\bibinfo  {journal} {J. Chem. Phys.}\ }\textbf {\bibinfo {volume} {103}},\
  \bibinfo {pages} {7422} (\bibinfo {year} {1995})}\BibitemShut {NoStop}%
\bibitem [{\citenamefont {Schultz}(1999)}]{schultz_local_1999}%
  \BibitemOpen
  \bibfield  {author} {\bibinfo {author} {\bibfnamefont {P.~A.}\ \bibnamefont
  {Schultz}},\ }\href {\doibase 10.1103/PhysRevB.60.1551} {\bibfield  {journal}
  {\bibinfo  {journal} {Phys. Rev. B}\ }\textbf {\bibinfo {volume} {60}},\
  \bibinfo {pages} {1551} (\bibinfo {year} {1999})}\BibitemShut {NoStop}%
\bibitem [{\citenamefont {Truflandier}\ and\ \citenamefont
  {Bowler}(2012)}]{truflandier__2012}%
  \BibitemOpen
  \bibfield  {author} {\bibinfo {author} {\bibfnamefont {L.~A.}\ \bibnamefont
  {Truflandier}}\ and\ \bibinfo {author} {\bibfnamefont {D.~R.}\ \bibnamefont
  {Bowler}},\ }\href@noop {} {\bibfield  {journal} {\bibinfo  {journal}
  {unpublished results}\ } (\bibinfo {year} {2012})}\BibitemShut {NoStop}%
\bibitem [{Note1()}]{Note1}%
  \BibitemOpen
  \bibinfo {note} {There is no restriction for the generation of the sampling
  points and the corresponding weights. It is well known that other nonlinear
  distributions allow to accelerate the convergence of numerical integration
  with respect to the size of the grid, mainly when we have to deal with
  singularities due to electronic cusps and nodal properties (for all--electron
  approaches) or Coulomb potentials. In this case, if one want to make an
  efficient use of FFT for the computation of Coulomb potential, we have to
  introduced transformation matrix to map the nonlinear distributions onto the
  FFT grid. We let this possibility open for future
  investigations.}\BibitemShut {Stop}%
\bibitem [{\citenamefont {Skylaris}\ \emph {et~al.}(2001)\citenamefont
  {Skylaris}, \citenamefont {Mostofi}, \citenamefont {Haynes}, \citenamefont
  {Pickard},\ and\ \citenamefont {Payne}}]{skylaris_accurate_2001}%
  \BibitemOpen
  \bibfield  {author} {\bibinfo {author} {\bibfnamefont {C.}~\bibnamefont
  {Skylaris}}, \bibinfo {author} {\bibfnamefont {A.~A.}\ \bibnamefont
  {Mostofi}}, \bibinfo {author} {\bibfnamefont {P.~D.}\ \bibnamefont {Haynes}},
  \bibinfo {author} {\bibfnamefont {C.~J.}\ \bibnamefont {Pickard}}, \ and\
  \bibinfo {author} {\bibfnamefont {M.~C.}\ \bibnamefont {Payne}},\ }\href
  {\doibase 10.1016/S0010-4655(01)00248-X} {\bibfield  {journal} {\bibinfo
  {journal} {Comput. Phys. Comm.}\ }\textbf {\bibinfo {volume} {140}},\
  \bibinfo {pages} {315} (\bibinfo {year} {2001})}\BibitemShut {NoStop}%
\bibitem [{\citenamefont {Kohn}(1996)}]{kohn_density_1996}%
  \BibitemOpen
  \bibfield  {author} {\bibinfo {author} {\bibfnamefont {W.}~\bibnamefont
  {Kohn}},\ }\href {\doibase 10.1103/PhysRevLett.76.3168} {\bibfield  {journal}
  {\bibinfo  {journal} {Physical Review Letters}\ }\textbf {\bibinfo {volume}
  {76}},\ \bibinfo {pages} {3168} (\bibinfo {year} {1996})}\BibitemShut
  {NoStop}%
\bibitem [{\citenamefont {Hern{\'a}ndez}\ \emph {et~al.}(1996)\citenamefont
  {Hern{\'a}ndez}, \citenamefont {Gillan},\ and\ \citenamefont
  {Goringe}}]{hernandez_linear-scaling_1996}%
  \BibitemOpen
  \bibfield  {author} {\bibinfo {author} {\bibfnamefont {E.}~\bibnamefont
  {Hern{\'a}ndez}}, \bibinfo {author} {\bibfnamefont {M.~J.}\ \bibnamefont
  {Gillan}}, \ and\ \bibinfo {author} {\bibfnamefont {C.~M.}\ \bibnamefont
  {Goringe}},\ }\href {\doibase 10.1103/PhysRevB.53.7147} {\bibfield  {journal}
  {\bibinfo  {journal} {Physical Review B}\ }\textbf {\bibinfo {volume} {53}},\
  \bibinfo {pages} {7147} (\bibinfo {year} {1996})}\BibitemShut {NoStop}%
\end{thebibliography}

\end{document}